\documentclass[twocolumn,prl,amsmath,amssymb]{revtex4}
 
\usepackage{graphicx}
\usepackage{dcolumn}
\usepackage{bm}
 
\newcommand{\kbf}{{\mathbf{k}}}
\newcommand{\dd}{{\mathrm{d}}}
\newcommand{\ee}{{\mathrm{e}}}
\newcommand{\Vdp}{V_{\mathrm{dp}}}
\newcommand{\perm}{{\mathrm{per}}}
\newcommand{\calD}{{\mathcal{D}}}
\newcommand{\calR}{{\mathcal{R}}}
\newcommand{\tT}{{\tilde{T}}}

\newcommand{\pvee}{{\scriptstyle\vee}}

\begin{document}
 
\preprint{APS/123-QED}
 
\title{Quantum fields and quantum groups}
 
\author{Christian
Brouder$^1$}\email{brouder@lmcp.jussieu.fr} 
\author{Robert Oeckl$^2$}\email{oeckl@cpt.univ-mrs.fr}
\affiliation{%
$^1$Laboratoire de Min\'eralogie Cristallographie, CNRS UMR 7590, 
Universit\'es Paris 6 et 7, IPGP, case 115, 4 place Jussieu, 
75252 Paris cedex 05, France\\
$^2$Centre de Physique Th\'eorique,
CNRS Luminy, F-13288 Marseille Cedex 9, France
}%
 
\date{\today}
 
\begin{abstract}
Quantum fields are shown to be an example of infinite-dimensional
quantum groups. A dictionary is established between quantum field
and quantum group concepts: the expectation value over the vacuum is
the counit, operator and time-ordered products are twisted products,
Wick's theorem is the definition of a twisted product.
Renormalization becomes the replacement of a coquasitriangular
structure by a 2-coboundary.
Quantum group methods are used to discuss
quantum field theory with a correlated initial state.
\end{abstract}
 
\maketitle
 
{\sl{Introduction}} --
Quantum groups are a powerful generalization of 
groups and Lie algebras \cite{Majid}. Since their
creation in the mid-eighties, quantum groups have 
attracted considerable attention
and more than four thousand papers were published 
on the subject in the mathematical literature. 
In physics, groups and Lie algebras are everywhere
and it can be expected that quantum groups
are useful too. Indeed, quantum groups play a role
in solid-state physics (magnetic systems
\cite{Wiegmann,Alcaraz,MartinDelgado}, phonons \cite{Bonechi})
which is directly linked to their origin in integrable systems
\cite{Korepin}.
Furthermore, they are the basic symmetry of the quantum Hall effect
\cite{Kogan,Grensing}.
In this paper we first show that quantum groups underlie
also a most successful and pervading tool of physics:
quantum field theory. Then we use quantum group
methods to solve a problem of quantum field
theory with a correlated initial state, to 
interpret renormalization and to explain the
origin of quasifree states.
 
{\sl{Definitions}} --
A quantum group is a Hopf algebra \cite{Swe:hopfalg} equipped with a 
quasitriangular structure \cite{Kassel,Majid}.
A Hopf algebra $H$ is an algebra: a vector space $H$
with an associative product denoted by $\cdot$ and
a unit $1$. It is also a coalgebra: there is a linear
map $\Delta : H\rightarrow H\otimes H$, called a coproduct,
which ``splits'' an element $a$ of $H$ into the sum of all its
left and right parts: $\Delta a = \sum a_{(1)}\otimes a_{(2)}$.
Coassociativity means that applying the coproduct again on the left
or the right hand component yields the same result.
In other words, $\Delta$ is coassociative if and only if
$ \sum (\Delta a_{(1)})\otimes a_{(2)} = 
\sum a_{(1)}\otimes (\Delta a_{(2)})$. In a coalgebra,
the coproduct is assumed coassociative.
A coalgebra is
also equipped with a counit $\varepsilon$, which is a map
from $H$ to $\mathbb{C}$ such that 
$\sum\varepsilon(a_{(1)}) a_{(2)}=\sum a_{(1)}\varepsilon(a_{(2)})=a$.
In a Hopf algebra, the coalgebra structure is compatible
with the algebra structure:
$\Delta (a\cdot b) = 
\sum (a_{(1)}\cdot  b_{(1)})\otimes (a_{(2)}\cdot b_{(2)})$
and $\varepsilon(a\cdot b)=\varepsilon(a)\varepsilon(b)$.
The final ingredient of a Hopf algebra is the antipode,
which is a map $S: H\rightarrow H$ such that
$\sum S(a_{(1)})\cdot a_{(2)}
= \sum a_{(1)}\cdot S(a_{(2)})=\varepsilon(a) 1$.
A Hopf algebra is commutative if the product $\cdot$
is commutative, and cocommutative if
$\sum a_{(1)}\otimes a_{(2)}=\sum a_{(2)}\otimes a_{(1)}$.
A Hopf algebra is in some sense a self-dual structure,
because the dual of a product is a coproduct and the
dual of a unit is a counit, and vice-versa.
 
To complete the definitions, a quantum group is a Hopf 
algebra together with an
element $R$ of $H\otimes H$ called a quasitriangular
structure or a universal $R$-matrix
\cite{Kassel}. Because of the self-dual nature of Hopf
algebras, quantum
group can also be described by the dual of a universal
$R$-matrix, called a coquasitriangular structure \cite{Majid}. 
This is the concept we need in quantum field theory.
A coquasitriangular structure is an invertible bilinear map $\calR: 
H\times H\rightarrow \mathbb{C}$ such that 
\begin{eqnarray}
\calR(a\cdot b,c) &=& \sum \calR(a,c_{(1)}) \calR(b,c_{(2)}),
\label{Laplace1}\\
\calR(a, b\cdot c) &=& \sum \calR(a_{(1)},c) \calR(a_{(2)},b).
\label{Laplace2}
\end{eqnarray}
For a commutative and cocommutative Hopf algebra, 
no other condition is required for $\calR$.
We restrict ourselves to this case in the following.
 
We use $\calR$ to define a twisted product,
denoted by $\circ$,
\begin{eqnarray}
a \circ b &=& \sum \calR(a_{(1)},b_{(1)}) \, a_{(2)}\cdot b_{(2)}.
\label{circle}
\end{eqnarray}
The twisted product is  a special case of
Sweedler's crossed product \cite{Sweedler}.
It can be checked that, when Eqs.(\ref{Laplace1}) and (\ref{Laplace2})
are satisfied and the coproduct is cocommutative, then
the twisted product is associative \cite{Sweedler}.
 
The quantum group stage is now set, and we show that quantum
field theory plays naturally on it. We take the 
example of a real scalar particle. The field operator is
\begin{eqnarray*}
\phi(x) &=& \int \frac{\dd\kbf}{(2\pi)^3\sqrt{2\omega_k}}
\Big(\ee^{-i p\cdot x} a(\kbf) + \ee^{i p\cdot x} a^\dagger(\kbf)\Big),
\end{eqnarray*}
where $\omega_k=\sqrt{m^2+|\kbf|^2}$, $p=(\omega_k,\kbf)$,
$a^\dagger(\kbf)$ and $a(\kbf)$ are the creation and annihilation operators
acting on $F_s$, the symmetric Fock space of scalar particles
\cite{ReedSimonII}.
To avoid ill-defined integrals, we consider the vector
space of smoothed fields
$V = \{\phi(f), f\in \calD(\mathbb{R}^{4})\}$,
where $\phi(f)=\int \dd x f(x) \phi(x)$ and 
$\calD(\mathbb{R}^{4})$ is the space of smooth functions on $\mathbb{R}^{4}$
with compact support \cite{ReedSimonII}.
 
{\sl{The quantum group of quantum fields}} --
Now we establish a connection between
the quantum group and quantum field concepts.
From the vector space $V$, we build the symmetric Hopf algebra
$S(V)=\bigoplus_{n=0}^\infty S^n(V)$ (see ref. \cite{Dascalescu}),
where $S^0(V)=\mathbb{C} 1$, $S^1(V)=V$ and $S^n(V)$ is generated
by the symmetric product of $n$ elements of $V$. An element of
$S^n(V)$ is said to have degree $n$.
The symmetric product in $S(V)$ is denoted by $\pvee$ and is induced
by the normal product of operators $a(\kbf)$ and $a^\dagger(\kbf)$:
in the usual quantum field notation,
$\phi(f)\pvee\phi(g)= {:}\phi(f)\phi(g){:}$.
Since  $\pvee$ is associative and commutative, $S(V)$ is a
commutative algebra with unit $1$, where $1$ is the 
unit operator acting on $F_s$.
The coproduct is defined on $V$ by
$\Delta \phi(f)=\phi(f)\otimes 1 + 1\otimes \phi(f)$,
and extended to $S(V)$ by compatibility with the product.
For example,
$\Delta (\phi(f)\pvee \phi(g)) =(\phi(f)\pvee\phi(g))\otimes 1 + 
\phi(f)\otimes\phi(g)+\phi(g)\otimes\phi(f)+
1\otimes (\phi(f)\pvee\phi(g))$. Note that this coproduct is 
cocommutative.  The counit is 
$\varepsilon(a)=0$ if $a\in S^n(V)$ with $n>0$ and
$\varepsilon(1)=1$. In the quantum field case, 
we can prove the additional identity
$\varepsilon(a)=\langle 0|a|0\rangle$,
where $a$ is any element of $S(V)$ and
$|0\rangle$ is the vacuum \cite{BrouderQFA}.
Finally, the antipode is $S(a)=(-1)^n a$ for 
$a\in S^n(V)$. This makes $S(V)$ a commutative and
cocommutative Hopf algebra.
It is even a Hopf $\star$-algebra, with the involution
$\phi(f)^\star=\phi(f^*)$, where $f^*$ the
complex conjugate of $f$.
 
To make $S(V)$ a quantum group, we must add a coquasitriangular
structure. This structure is entirely determined by its value
on $V$. More precisely, if
$a=u_1\pvee \dots \pvee u_m \in S^m(V)$
and
$b=v_1\pvee \dots \pvee v_n \in S^n(V)$,
where $u_i$ and $v_j$ are elements of $V$, 
then Eqs.(\ref{Laplace1}) and (\ref{Laplace2})
imply that
$\calR(a,b)=0$ if $m\not=n$ and
$\calR(a,b)=\perm(u_i,v_j)$ if $m=n$ \cite{Grosshans}.
The permanent $\perm(u_i,v_j)$ is a kind of determinant
of the matrix $\calR(u_i,v_j)$ without minus signs:
\begin{eqnarray}
\perm(u_i,v_j) &=& \sum_{\sigma} \calR(u_1,v_{\sigma(1)}) \cdots
\calR(u_n,v_{\sigma(n)}),\label{permanent}
\end{eqnarray}
where the sum is over all permutations $\sigma$ of $\{1,\dots,n\}$. 
Therefore, the most general Poincar\'e-invariant 
coquasitriangular structure
on $S(V)$ is determined by its value on $V$, which can be written
\begin{eqnarray}
\calR(\phi(f),\phi(g)) &=& \int \dd x \dd y f(x) G(x-y) g(y),
\end{eqnarray}
where $G(x)$ is a Lorentz invariant distribution.
As explained above, $\calR$ gives rise to a twisted product on $S(V)$
defined by (\ref{circle}).  By choosing the proper $G(x)$,
this twisted product can become the operator product or the 
time-ordered product. Explicitly, if 
\begin{eqnarray*}
G(x)&=&G_+(x) = \int \frac{d\kbf}{(2\pi)^3 2\omega_k} \ee^{-ip\cdot x},
\end{eqnarray*}
the twisted product defined by (\ref{circle}) is the
usual product of field operators in quantum field theory.
If $G(x)$ is the Feynman propagator $G_F(x)$, with
\begin{eqnarray*}
G_F(x) &=& 
i\int \frac{d^4p}{(2\pi)^4} \frac{\ee^{-ip\cdot x}}{p^2-m^2+i\epsilon},
\end{eqnarray*}
then the twisted product defined by (\ref{circle}) is
the time-ordered product. 
Let us show this for the operator product.
The twisted product of two fields is
$\phi(f)\circ\phi(g)=\phi(f)\pvee\phi(g) + \calR(\phi(f),\phi(g))$
and, by Wick's theorem, the operator product of two
fields is
$\phi(f)\phi(g)=\phi(f)\pvee\phi(g) + 
\langle 0 | \phi(f)\phi(g) |0\rangle$. Thus, the two
products are identical if we define
$\calR(\phi(f),\phi(g))=\langle 0 | \phi(f)\phi(g) |0\rangle$.
This identification gives us $G(x)=G_+(x)$.
Once the products are identified for two fields,
they can be identified for any number of fields
by noticing that Eq.(\ref{circle}) is Wick's theorem 
under a quantum group disguise \cite{BrouderQFA}:
Eq.(\ref{circle}) can be described in words 
by ``the twisted product of two normal products
$a$ and $b$ is obtained by summing all possible
pairings between $a$ and $b$'', which is 
Wick's theorem for the product of $a$ and $b$
\cite{Wick}, the pairing being defined
by $\langle 0 | \phi(f)\phi(g)
|0\rangle=\calR(\phi(f),\phi(g))$.
Since Wick's theorem is valid for operator products
and time-ordered products \cite{Gross},
the twisted product is equal to the
time-ordered product if we identify
$\calR(\phi(f),\phi(g))$
with $\langle 0 | T\phi(f)\phi(g) |0\rangle$.
This leads to $G(x)=G_F(x)$. Note that
this type of relation between Wick's theorem and Hopf algebras
was already pointed out by Fauser \cite{Fauser}.
 
In this context, the normal product becomes the basic product
of quantum field theory, while the operator product and the
time-ordered product become deformed products. The advantage
of this point of view was already stressed in the fifties
by Gupta \cite{Gupta} and is the basis of the star-product
approach to quantum field theory \cite{Dito92,Dutsch}.
 
{\sl{Applications}} --
We have shown that many concepts and tools of quantum field theory,
such as normal product, expectation value over the vacuum,
Wick's theorem, operator and time-ordered products,
are natural elements of the quantum group point of view.
Now we show that quantum group methods can solve elegantly
quantum field theory problems.  As a simple example, consider a system 
with a correlated initial state $|\Omega\rangle$. In solid-state and
atomic physics, the initial state is usually described by a Slater
determinant \cite{Fetter}.
However, in the case of degenerate initial states
\cite{Esterling,Kuo,Cederbaumarticle}, or when discussing 
non-equilibrium Green functions \cite{Hall,Henning,FauserWolter}
or contracted Schr\"odinger equations
(see \cite{Mukherjee,MukherjeeErratum,Mazziotti} and
references therein), a correlated initial state $|\Omega\rangle$
must be used. The main difficulty of this problem comes
from the fact that the expectation value of
normal products over $|\Omega\rangle$ is not zero,
so that standard quantum field methods cannot be used.
To solve this difficulty, Kutzelnigg and Mukherjee \cite{Kutzelnigg}
proposed to define a normal product adapted to $|\Omega\rangle$:
a transformation $a\rightarrow \tilde a$ such that
$\langle \Omega| \tilde a |\Omega\rangle=\varepsilon(a)$.
We can generalize this question by taking a linear map
$\omega$ from $S(V)$ to $\mathbb{C}$ such that $\omega(1)=1$
and looking for a transformation $a\rightarrow \tilde a$ such that
$\omega(\tilde a)=\varepsilon(a)$. The original problem
is recovered whith $\omega(a)=\langle \Omega| a |\Omega\rangle$,
but $\omega$ can also represent a mixed initial state.
This problem was solved explicitly for elements
$a$ of small degrees in \cite{Kutzelnigg} but the 
general solution can now be obtained by quantum group methods. 
We call 1-cochain a linear map $\rho$ from
$S(V)$ to $\mathbb{C}$ such that $\rho(1)=1$.
In particular, $\omega$ is a 1-cochain.
In a foundational paper on Hopf algebra cohomology \cite{Sweedler},
Sweedler showed that  
two 1-cochains $\omega$ and $\rho$ can be multiplied
by defining $(\omega\star\rho)(a)
=\sum \omega(a_{(1)})\rho(a_{(2)})$. The set of 
1-cochains forms an Abelian group with unit element $\varepsilon$.
If $\omega^{-1}$ is the unique 1-cochain defined
by $\omega\star\omega^{-1}=\varepsilon$, then
the normal products adapted to $\omega$ are obtained
by defining
$\tilde a = \sum \omega^{-1}(a_{(1)}) a_{(2)}$.
This solves the problem because
$\omega(\tilde a)= (\omega^{-1}\star\omega)(a)=\epsilon(a)$.
If we compare this general solution with the first
degrees given in \cite{Kutzelnigg}, we observe that 
the quantum group method tames the combinatorial complexity
of the problem and provides a way to manipulate the
general term of the solution.
 
We saw that quantum groups fare well for the free scalar field
and the extension to fermions is done by replacing
Hopf algebras by Hopf superalgebras \cite{BrouderQG}. 
But what happens with 
interacting fields? We must extend our vector
space $V$ to include normal products of fields at a point
$\Vdp = \{\phi^{(n)}(f), f\in \calD(\mathbb{R}^{4})\}$,
where $\phi^{(n)}(f)=\int \dd x f(x) \phi^{(n)}(x)$ and 
$\phi^{(n)}(x)$ is the divided power of $n$ fields
at $x$:
$\phi^{(n)}(x)=(1/n!) \phi(x)\pvee\dots\pvee\phi(x)$.
We choose divided powers instead of usual powers of fields because
their coproduct is simpler:
\begin{eqnarray}
\Delta \phi^{(n)}(x) &=& \sum_{k=0}^n \phi^{(k)}(x)
\otimes \phi^{(n-k)}(x).
\label{divided}
\end{eqnarray}
This coproduct is now extended to $S(\Vdp)$
by using $\Delta (a\pvee b)=(\Delta a)
(\pvee\otimes\pvee)(\Delta b)$.
The coquasitriangular structure of two divided powers
is calculated from
Eqs.(\ref{Laplace1}) and (\ref{Laplace2}):
\begin{eqnarray}
\calR(\phi^{(m)}(x),\phi^{(n)}(y))
&=& \frac{\delta_{m,n}}{n!} G(x-y)^n.
\end{eqnarray}
In general, $G(x)$ is a distribution and its power
is not well defined. Using microlocal arguments, H\"ormander
proved that distributions can be multiplied if their
wavefront sets satisfy a compatibility condition
\cite{ReedSimonII}.
It turns out that $G_+(x)$ satisfies this condition, so
that the operator product of powers of fields can be defined.
In fact, Brunetti and Fredenhagen \cite{Brunetti}
argued that, up to smooth functions, $G_+(x)$ is
the only distribution that solves the Klein-Gordon equation,
that defines a state, and 
whose powers $G_+^n(x)$ exist as distributions.
On the other hand, the Feynman propagator $G_F(x)$ does not 
satisfy the wavefront set condition
and $G_F(x)^n$ is not defined. 
This is the origin of the renormalization problem. 
In fact, no coquasitriangular structure can reproduce
the renormalized time-ordered product, and the
quantum group approach seems to fail. 
 
However, quantum group theory is quite resourceful. To see how it
can be used to deal with renormalization, we use the Epstein-Glaser
approach, which is a renormalization
method without infinities.
Epstein and Glaser \cite{Epstein} define a linear
map $T$ from $S(\Vdp)$ to $S(\Vdp)$, that we call
the $T$-map (they call it $T$-products but we prefer
$T$-map to avoid confusion).
This map satisfies the equations $T(1)=1$,
$T(u)=u$ for $u\in\Vdp$ and
$T(a)=\sum \varepsilon\big(T(a_{(1)})\big) a_{(2)}$
for any $a\in \Vdp$.
In these equations, which are also valid in curved space-times
\cite{Brunetti}, $T(a)$ is renormalized (i.e.\ gives rise to finite
quantities).
As a side remark, the efficiency of the quantum group
language can be appreciated by noting that 
our equation for $T(a)$ corresponds to a
two-line formula in the original papers
(Eq.(42) in \cite{Epstein}, Eq.(26) in \cite{Brunetti}).
The importance of the $T$-map stems from the fact that
it enables us to calculate the renormalized S-matrix.
For an interacting Lagrangian $\lambda \phi^{(n)}(f)$,
the renormalized S-matrix is \cite{Scharf}
\begin{eqnarray*}
S(f) &=& T\Big(1 + \sum_{k=1}^\infty \frac{(-i\lambda)^k}{k!}
\phi^{(n)}(f)\pvee\cdots\pvee\phi^{(n)}(f)\Big).
\end{eqnarray*}
 
Let us investigate what quantum groups can tell us about the $T$-map.
The function $t(a)=\varepsilon\big(T(a)\big)$ is a
linear map from $S(\Vdp)$ to $\mathbb{C}$ such that
$t(1)=1$. Thus, it is a 1-cochain. For any 1-cochain
$\rho$, Sweedler defines the 2-cocycle
$(\partial\rho)(a,b)=\sum \rho(a_{(1)})\rho(b_{(1)})
\rho^{-1}(a_{(2)}\pvee b_{(2)})$.
In particular, we call $\chi=\partial t^{-1}$ the
2-cocycle obtained from the inverse of $t$.
In ref.\cite{Sweedler}, Sweedler shows that any 
2-cocycle $\chi$ can be used to define an associative
product 
$a\circ b=\sum \chi(a_{(1)},b_{(1)}) a_{(2)}\pvee b_{(2)}$.
Now we define the map $\tT$ from
$S(\Vdp)$ to $S(\Vdp)$ by
$\tT(1)=1$, $\tT(u)=u$ if $u\in\Vdp$ and
$\tT(a\pvee b)=\tT(a)\circ\tT(b)$,
so that
$\tT(u_1\pvee\dots\pvee u_n)=u_1\circ\dots\circ u_n$
for $u_i$ in $\Vdp$. Then it can be shown that $\tT=T$. 
This fact has several consequences.
From the quantum field point of view, it shows that
the $T$-map arises from a renormalized product $\circ$
which is associative. This fact seems not to have been
realized before.
From the quantum group point of view, 
renormalization is the replacement of
a coquasitriangular structure
by a more general 2-cocycle. However, not all 2-cocycles play a role in
renormalization. The
set of ``renormalizing" 2-cocycles should be
deduced from the Connes-Kreimer
Hopf algebra of renormalization \cite{CKI,CKII,PinterHopf}.
 
Let us give two last examples of the contact between quantum 
fields and quantum groups. We saw that any 
1-cochain gives rise to a 2-cocycle. 
Since coquasitriangular structures are
2-cocycles, we could
wonder which 1-cochains $\omega$ give rise to
a coquasitriangular structure $\calR$.
The answer is very simple: the 1-cocyles 
that give rise to $\calR$
can be defined recursively by
$\rho(a\vee b)=\sum \rho(a_{(1)})\rho(b_{(1)})
\calR(a_{(2)},b_{(2)})$.
If $\calR(u^\star,v^\star)=\calR(v,u)^*$, $\calR(u^\star,u)\ge0$
and $\rho(u)=0$ for any $u,v$ in $V$, then
$\rho$ is a state for $S(V)$ (i.e.\
$\rho(a^*\vee a)\ge0$ for any $a\in S(V)$).
But these states are very special. They are called
quasifree states.
These states are fundamental to define a quantum
field theory in curved spacetimes \cite{Kay1,Kay2}.
In fact, the singular structure of any physically 
reasonable state of a scalar field in curved spacetimes
is determined by a quasifree state \cite{Hollands3}.
The concept and importance of quasifree states emerged
slowly in quantum field theory. Their peculiar
nature is immediately clear from the quantum group
point of view: they are the only states
that derive from a coquasitriangular structure.
Finally, we consider a last problem posed by
Kutzelnigg and Mukherjee \cite{Kutzelnigg}.
When the normal products $\tilde a$
adapted to $\omega$ are defined, how can we
calculate their operator products?
From the quantum group point of view, the
answer is straightforward. If
$\calR$ is the coquasitriangular structure
used to define the product $a\circ b$, then the result is
\begin{eqnarray*}
{\tilde a}\circ {\tilde b} &=&
\sum \chi(a_{(1)},b_{(1)})\, \widetilde{a_{(2)}\pvee b_{(2)}},
\end{eqnarray*}
where the 2-cocycle $\chi$ is
$\chi=(\partial \omega^{-1}) \star \calR$.
Without the quantum group structure in sight, the
same result can be obtained by hand for the first
degrees, but its combinatorial complexity is
appalling.
 
{\sl{Conclusions}} --
We showed that quantum groups provide a
fresh look at many aspects of quantum field theory
and they introduce powerful new concepts, such as
coquasitriangular structures and cochains.
On the practical side, quantum groups replace
combinatorial proofs by algebraic proofs.
On the conceptual side, quantum groups
provide a second quantization without commutators
that enables us to go beyond standard quantum field theory.
For example, it allows the extension of quantum field theory to
systems with quantum group symmetries
\cite{Oeckl}.
That work employs a path integral approach
which is dual (in the sense of quantum groups) to the present
approach. (For comparison note that our $\calR$ is encoded there in
$\gamma$.)
In particular, this can be used to describe systems with anyonic
particles
\cite{Oe:spinstat} or models with noncommutative space-time structures
\cite{Oe:nctwist}. 
 
\begin{acknowledgments}
We thank A. Frabetti, J.-L. Loday and B. Fauser for
useful discussions, and M. Calandra and F. Mauri for their comments.
R.~O.\ was supported by a Marie Curie fellowship grant from the
European Union.
This is IPGP contribution \#0000.
\end{acknowledgments}


%
\end{document}